# Dynamic critical behavio(u)r of a cluster algorithm for the Ashkin–Teller model

Jesús Salas[a] and Alan D. Sokal[a,*]

[a]Department of Physics, New York University, 4 Washington Place, New York, NY 10003 USA

We study the dynamic critical behavior of a Swendsen–Wang–type algorithm for the Ashkin–Teller model. We find that the Li–Sokal bound on the autocorrelation time ($\tau_{\mathrm{int},\mathcal{E}} \geq \mathrm{const} \times C_H$) holds along the self-dual curve of the symmetric Ashkin–Teller model, but this bound is apparently not sharp. The ratio $\tau_{\mathrm{int},\mathcal{E}}/C_H$ appears to tend to infinity either as a logarithm or as a small power ($0.05 \lesssim p \lesssim 0.12$).

Monte Carlo simulations near critical points are hampered by critical slowing-down: the autocorrelation time $\tau$ grows like $L^z$ for a system of linear size $L$ at criticality. The traditional local algorithms have a dynamic critical exponent $z \gtrsim 2$. Cluster algorithms [1,2] can in some cases do much better: the Swendsen–Wang (SW) algorithm for the ferromagnetic Potts model [1] has $z$ between 0 and 1, depending on the number of states and the dimensionality of the lattice. But there is little *theoretical* understanding of the dynamic critical behavior of SW-type algorithms. We have only a rigorous *lower bound* [3]

$$\tau_{\mathrm{int},\mathcal{E}},\ \tau_{\mathrm{exp}} \ \geq \ \mathrm{const} \times C_H \tag{1}$$

where $C_H$ is the specific heat, and hence

$$z_{\mathrm{int},\mathcal{E}},\ z_{\mathrm{exp}} \ \geq \ \frac{\alpha}{\nu}\,, \tag{2}$$

where $\alpha$ and $\nu$ are the standard *static* critical exponents.

Obviously we would like to know whether these bounds are *sharp*: that is, does (2) holds as *equality* or as a *strict inequality*? Unfortunately, the numerical data for two-dimensional Potts models are not very conclusive. For the Ising model, the bound would be sharp if $\tau$ grows like a logarithm. However, the available data are compatible with an exponent $z \sim 0.23$ [1,4], with a logarithm [5], and even with a behavior $\tau \sim L^{1/8} \log L$ [4]. For the 3-state Potts model the bound seems to be *not sharp*: $z = 0.55 \pm 0.03$ [3] versus $\alpha/\nu = 2/5$. Finally, the 4-state Potts model is very special: a naive fit to the data gives $z = 0.89 \pm 0.05$ [3], which is smaller than $\alpha/\nu = 1$. This anomalous behavior can be explained if we take into account the true leading term of the specific heat $C_H \sim L \log^{-3/2} L$ [6]. This suggests that $\tau$ might have a similar multiplicative logarithmic term, in which case the bound would be sharp (modulo a possible logarithm).

There is yet another way of "interpolating" between the 2-state (Ising) and 4-state Potts models: both are particular cases of the Ashkin–Teller (AT) model [7]. As a matter of fact, the self-dual curve of the symmetric AT model joins the critical points of these two models (see Figure 1). Along this curve the static critical exponents vary continuously.

An SW-type algorithm for the AT model was first devised by Wiseman and Domany [8]. Here we study a simplified ("embedding") version of this algorithm: we want to know how the dynamic critical exponent $z$ behaves along the self-dual curve, and in particular whether the Li–Sokal bound (1)/(2) is sharp. We have performed simulations at three different points on the AT self-dual curve (see Table 1). In addition, we have reanalyzed the data obtained by Baillie and Coddington [4,9] for the Ising model at criticality. We conclude that the bound (1) is *almost but not quite sharp*: the ratio $\tau_{\mathrm{int},\mathcal{E}}/C_H$ tends to infinity as $L \to \infty$, either as a logarithm or as a small power ($0.05 \lesssim p \lesssim 0.12$).

The Ashkin–Teller (AT) model [7] is a general-

*Speaker at the conference.

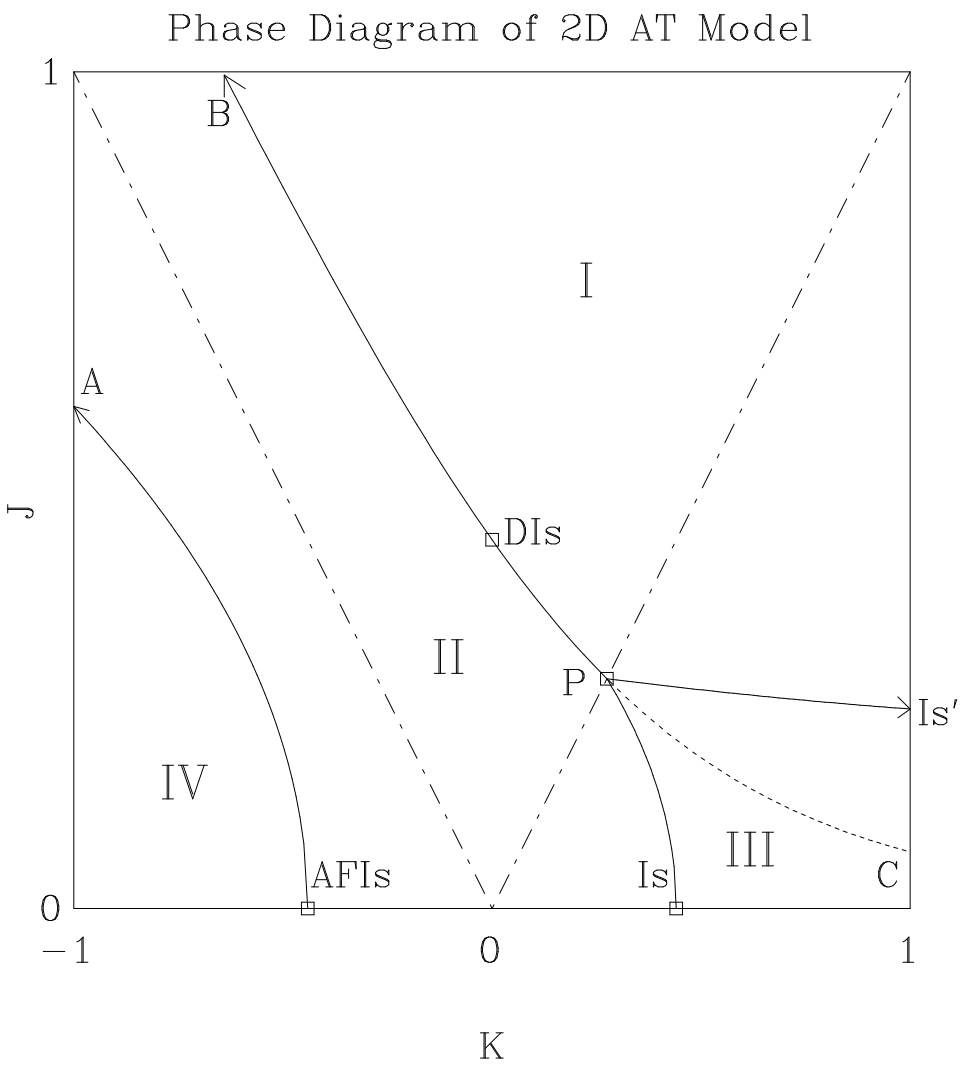

Figure 1. Phase diagram of the symmetric Ashkin-Teller model on the square lattice. The self-dual curve is B–DIs(decoupled Ising)–P(Potts)–C. The solid curves represent second-order phase transitions, and the dotted curve is the non-critical part of the self-dual curve. The dash-dotted curve is the 4-state Potts-model subspace. The roman numerals designate the different phases of the model (see text).

ization of the Ising model to a four-state model. To each lattice site $x$ we assign two Ising spins $\sigma_x = \pm 1$ and $\tau_x = \pm 1$, and they interact through the Hamiltonian

$$H = -J \sum_{\langle xy \rangle} \sigma_x \sigma_y - J' \sum_{\langle xy \rangle} \tau_x \tau_y - K \sum_{\langle xy \rangle} \sigma_x \tau_x \sigma_y \tau_y \ , \quad (3)$$

where the sums run over nearest-neighbor pairs $\langle xy \rangle$. Note that the fields $\sigma$, $\tau$ and $\sigma\tau$ play symmetric roles in this model; we can consider any two of these three as the “fundamental fields”. The plane $K = 0$ corresponds to a pair of decoupled Ising models with interactions $J$ and $J'$, while the line $J = J' = K$ is the 4-state Potts model with $J_{\rm Potts} = 4J$.

The family (3) of AT Hamiltonians exhibits several symmetries. First of all, we can permute freely the spin variables $(\sigma, \tau, \sigma\tau)$; this is equivalent to permuting the couplings $(J, J', K)$. Secondly, if the lattice is *bipartite* we can flip $\sigma$ or $\tau$ or both on one of the two sublattices; this is equivalent to flipping the sign on any *two* of the three couplings $(J, J', K)$.

| Point | $J = J'$ | $K$ |
|---|---|---|
| P (4-state Potts) | 0.274653 | 0.274653 |
| ZF | 0.302923 | 0.220343 |
| X2 | 0.344132 | 0.147920 |
| DIs (decoupled Ising) | 0.440687 | 0 |

Table 1
Points of the self-dual curve of the symmetric AT model where our MC simulations were performed. We also include the values corresponding to the Ising model (DIs).

In this paper we are interested in the 2D square-lattice symmetric $(J = J')$ AT model:

$$H_{\rm S} = -J \sum_{\langle xy \rangle} (\sigma_x \sigma_y + \tau_x \tau_y) - K \sum_{\langle xy \rangle} \sigma_x \tau_x \sigma_y \tau_y \ . \quad (4)$$

This model exhibits a rich phase diagram [10,11], which is shown in Figure 1. There are four different phases:

I) Baxter phase. The spins $\sigma$ and $\tau$ are independently ferromagnetically ordered.
II) Paramagnetic phase. Here the three spins $\sigma$, $\tau$ and $\sigma\tau$ are disordered.
III) The spins $\sigma$ and $\tau$ are disordered, but their product $\sigma\tau$ is ferromagnetically ordered.
IV) Both $\sigma$ and $\tau$ are disordered, but their product $\sigma\tau$ is antiferromagnetically ordered.

The curves separating these phases are critical. All except the curve B–P (i.e. the self-dual curve) are expected to be Ising-like, and their exact locations are unknown. The self-dual curve is given by [11]

$$e^{-2K} = \sinh 2J \ , \quad (5)$$

and the part above the 4-state Potts point P is critical. The critical exponents vary continuously along the self-dual curve, and their values are exactly known by relating the AT model with the Gaussian model [12].

Let us consider the general AT Hamiltonian (3) with the condition

$$J, J' \geq |K| \ . \quad (6)$$

(Actually, by using the symmetries, *any* AT model on a bipartite lattice can be mapped onto an equivalent model satisfying this condition.) One can then construct a SW-type algorithm for this model, following the general scheme of [13]: the idea is to decompose the Boltzmann weight associated with a given bond as a linear combination of Kronecker deltas of the spins, and then to introduce new auxiliary variables which live on the bonds. The final result is essentially the same algorithm as introduced by Wiseman and Domany [8]. Details of the derivation can be found elsewhere [14]. For the "multi-cluster" version of this algorithm, we have proven [14] the Li–Sokal bound (1)/(2) by following the scheme of the original proof [3].

A simpler SW-type algorithm can be introduced by considering the Boltzmann weight of a given bond $\langle xy \rangle$, conditional on the $\{\tau\}$ configuration: it is

$$W_{\rm bond}(\sigma_x, \sigma_y; \tau_x, \tau_y) = (1 - p_{xy}) + p_{xy}\delta_{\sigma_x,\sigma_y} \ , \quad (7)$$

where $p_{xy} = 1 - \exp(J + K\tau_x\tau_y)$. This system of $\sigma$ spins can be simulated using the standard SW algorithm with *ferromagnetic* effective nearest-neighbor coupling $J^{\rm eff}_{xy} = J + K\tau_x\tau_y$. An identical argument applies to the $\{\tau\}$ spins when the $\{\sigma\}$ spins are held fixed. The "embedding" algorithm for the AT model consists therefore of one SW update of the $\{\sigma\}$ spins with the $\{\tau\}$ spins held fixed, followed by SW update of the $\{\tau\}$ spins with the $\{\sigma\}$ spins held fixed.

We have simulated this embedding algorithm for the 4-state Potts model at criticality and for two other points (X2 and ZF) on the self-dual curve (see Table 1). X2 was already considered in [8], and ZF is the model introduced by Zamolodchikov and Fateev [15]. We have studied lattice sizes ranging from $L = 16$ to $L = 512$ (1024 for the 4-state Potts model). The number of measurements ranges from $8 \times 10^5$ to $4.4 \times 10^6$. (In units of the autocorrelation times, our run lengths are always at least $10^4\tau$, except for $L = 1024$ where we reached only $1500\tau$.) In all cases we discarded $10^5$ iterations for equilibration; this is always $\gtrsim 150\tau$. For the Potts case we have measured the energy, specific heat, second-moment correlation length and susceptibility. In the other two cases, we have measured these quantities for both the $\sigma$, $\tau$ spins and for the product $\sigma\tau$.

We have performed standard weighted least-squares fits to a power-law Ansatz for all the diverging static quantities (specific heat and susceptibilities). In these fits we use the data with $L \geq$ a cutoff $L_{min}$, and we vary $L_{min}$ until we obtain a good $\chi^2$ value; in this way we can protect against corrections to scaling. Our results (see Table 2) agree fairly well with the exact answers. The specific heat of the 4-state Potts model is the only exception. But in this case the true leading behavior is $\sim L \log^{-3/2} L$ [6], not merely $\sim L$. If we include this theoretical input in the fit we get a much better estimate [14].

We have also measured the autocorrelation functions and the corresponding integrated and exponential autocorrelation times (see e.g. [16]). Power-law fits yield the dynamic critical exponents $z_{{\rm int},\mathcal{E}}$ reported in Table 2. We see that the value of $z_{{\rm int},\mathcal{E}}$ is always slightly higher than the effective value of $\alpha/\nu$. Thus, the Li–Sokal bound (2) is satisfied — but apparently not as equality — along the AT self-dual curve (5).

Nevertheless, the very small values obtained here for $z_{{\rm int},\mathcal{E}} - \alpha/\nu$ (less than 0.11) suggest that the bound (1)/(2) might be sharp modulo a *logarithm*. To check this, we have studied the ratio $\tau_{{\rm int},\mathcal{E}}/C_H$. This ratio is in all cases an increasing function of $L$. We tried fits to ${\rm const} + AL^{-\Delta}$ and ${\rm const} + A/\log L$, but the results are poor: either $\chi^2$ is too large, or the value of the constant $A$ is implausibly large. We conclude that $\tau_{{\rm int},\mathcal{E}}/C_H$ probably does tend to infinity as $L \to \infty$. The question is: in what way? We next tried to fit $\tau_{{\rm int},\mathcal{E}}/C_H$ to a pure power law $AL^p$. The $\chi^2$ values are very good in all cases; the power $p$ is small and seems to increase from the Ising and X2 models ($p \approx 0.05$) to the 4-state Potts model ($p \approx 0.12$). If this is the true behavior, it would mean that the bound (1)/(2) fails to be sharp by only a small power. On the other hand, we have also tried the Ansatz $\tau_{{\rm int},\mathcal{E}}/C_H = A + B\log L$. The fits are again quite good for all the models. In this case the bound would fail to be sharp only by a multiplicative logarithm. It is very hard to distinguish numerically between these two scenarios: a logarithmic behavior can be quite well mimicked by a power law when the range of variation of $\log L$ is not very large. Indeed, we can equally well fit the data to a function $\log^p L$. Details of the data and the fits will be reported in [14].

| Ratio | 4-state Potts model | | ZF model | | X2 model | | Ising model | |
|---|---|---|---|---|---|---|---|---|
| | numerical | exact | numerical | exact | numerical | exact | numerical | exact |
| $\gamma/\nu$ | $1.744 \pm 0.001$ | 7/4 | $1.750 \pm 0.004$ | 7/4 | $1.751 \pm 0.001$ | 7/4 | | 7/4 |
| $\gamma'/\nu$ | $1.744 \pm 0.001$ | 7/4 | $1.668 \pm 0.005$ | 5/3 | $1.605 \pm 0.001$ | 1.6045 | | 1/2 |
| $\alpha/\nu$ | $0.768 \pm 0.009$ | $1 \times \log^{-3/2}$ | $0.663 \pm 0.006$ | 2/3 | $0.438 \pm 0.008$ | 0.4183 | | log |
| $z_{\mathrm{int},\mathcal{E}}$ | $0.876 \pm 0.012$ | $\geq 1 \times \log^{-3/2}$ | $0.740 \pm 0.010$ | $\geq 2/3$ | $0.477 \pm 0.028$ | $\geq 0.4183$ | $0.240 \pm 0.004$ | $\geq$ log |

Table 2
Static critical exponents and dynamic critical exponent ($z_{\mathrm{int},\mathcal{E}}$) coming from power-law fits to the Monte Carlo results [14]. For the Ising model we include our own fits to the dynamical data reported in Refs. [4,9]. Errors are one standard deviation. The symbol $1 \times \log^{-3/2}$ (resp. log) means that the leading term of the specific heat for the 4-state Potts model (resp. the Ising model) behaves like $L \log^{-3/2} L$ (resp. $\log L$).

In summary, the fits of the ratio $\tau_{\mathrm{int},\mathcal{E}}/C_H$ suggest two different scenarios

$$\tau_{\mathrm{int},\mathcal{E}}/C_H = \begin{cases} AL^p & \text{with } 0.05 \lesssim p \lesssim 0.12 \\ A + B \log L \end{cases} \qquad (8)$$

To distinguish between these two behaviors would require high-precision data on significantly larger lattices than those simulated here, probably up to at least $L = 2048$.

These computations were carried out at NYU, the Pittsburgh Supercomputing Center, and the Cornell Theory Center. The authors' research was supported in part by a M.E.C. (Spain)/Fulbright fellowship (J.S.), and by NSF grant DMS–9200719 (J.S. and A.D.S.).